\documentstyle[12pt,aaspp4]{article}
\def\h{Hipparcos }

\def\p{parallax }
\def\be{\begin{equation}}
\def\ee{\end{equation}}

\begin{document}

{\title{A PRECISION TEST OF HIPPARCOS SYSTEMATICS TOWARDS THE HYADES}
\author{\bf 
Vijay K. Narayanan and
Andrew Gould \footnote{Alfred P.\ Sloan Foundation Fellow}
}
\affil{Department of Astronomy, The Ohio State University, Columbus, OH 43210;}
\affil{ Email: vijay,gould@astronomy.ohio-state.edu}

\bigskip
\bigskip
\bigskip
\centerline{\bf ABSTRACT}
\medskip

We propose a test that can in principle detect any systematic errors in the 
Hipparcos parallaxes towards the Hyades cluster at the level of 0.3 mas.
We show that the statistical parallax algorithm subsumes the classical moving
cluster methods and provides more precise estimates of the distance
and the first two moments of the velocity distribution of the Hyades 
cluster namely, its bulk space velocity and the velocity dispersion tensor.
To test the \h parallaxes, we first rescale the bulk velocity determined
from statistical parallax to force agreement with the distance scale
determined from \h parallaxes.
We then predict the parallaxes of Hyades cluster members using this common 
cluster space velocity and their Hipparcos proper motions.
We show that the parallaxes determined in this manner ($\pi_{\rm pm}$) are 
consistent at the $1\sigma$ level with the parallaxes ($\pi_{\rm orb}$) of 
three Hyades spectroscopic binary systems with orbital solutions.
We find that $\left< \pi_{\rm pm} - \pi_{\rm orb} \right> = 0.52 \pm 0.47$ mas,
where the error is dominated by the errors in the orbital parallaxes.
A reduction in these errors would allow a test of the systematic errors in the 
Hipparcos parallaxes at the 0.3 mas level.
If the Hyades distance scale is fixed by \h parallaxes, then its bulk
velocity in equatorial coordinates is
$(V_{x}, V_{y}, V_{z}) = (-5.70 \pm 0.20, 45.62 \pm 0.11, 5.65 \pm 0.08)$ ${\rm \ km}\, {\rm s}^{-1}$,
its velocity dispersion is $320 \pm 39$ ${\rm m}\, {\rm s}^{-1}$, and the
distance modulus to the centroid of our sample of $43$ cluster members 
is $3.34 \pm 0.02$ mag.


\section{INTRODUCTION}

The Hipparcos mission (\cite{esa97}) has derived accurate astrometric 
parameters for about 120,000 stars distributed all over the sky.
The systematic errors in the \h parallaxes are estimated
to be $ \la 0.1 $ mas, while the random errors are
of the order of 1 mas (\cite{arenou95}; \cite{arenou97}).
However, recent comparisons of the distances to open clusters derived
from \h parallaxes and main sequence fitting techniques show surprisingly
large differences for some clusters (\cite{mermio97hip}; \cite{robichon97}),
which can be reconciled if the systematic error in the \h parallaxes
 is at the level of 1 mas, at least on small angular scales
(Pinsonneault et al. 1998, hereafter PSSKH98).
It is therefore prudent to compare the Hipparcos parallaxes with accurate
parallaxes determined in an independent manner.
In this paper, we propose and apply one such method which could in principle
test for systematic errors in Hipparcos astrometry towards the Hyades cluster
at the level of 0.28 mas.

The statistical power of our test arises from the fact that the 
fractional error in the Hipparcos proper motions of the Hyades cluster 
members ($\sigma_{\mu}/\left< \mu_{\rm Hya} \right> = 1.4 \%$) is about 
four times smaller than the fractional error  in their Hipparcos 
parallaxes ($\sigma_{\pi}/\left< \pi_{\rm Hya} \right> = 6 \%$), where
we have assumed that the mean parallax and the proper motion of the Hyades 
cluster are $\left< \pi_{\rm Hya} \right> = 21.5$ mas and 
$\left< \mu_{\rm Hya} \right> = 111$ mas$\,$yr$^{-1}$ respectively and 
their errors are $\sigma_{\pi} = 1.3$  mas and $\sigma_{\mu} = 1.5$ 
mas$\,$yr$^{-1}$.
Hence, if one can accurately determine the space velocity of the Hyades
cluster, one can use the Hipparcos proper motions to predict the parallaxes
of the individual Hyades members more accurately than the Hipparcos 
parallaxes, assuming that all the members partake in this common cluster 
motion (to within the velocity dispersion of the cluster).
The accurate parallaxes predicted in this manner can then be compared 
with the Hipparcos parallaxes and with parallaxes determined in an independent
manner to determine the level of the systematic errors in the Hipparcos
astrometry.

The basic steps of our test are as follows:
\begin{description}
\item [{(1)}:] We select a set of Hyades cluster members all of which 
are consistent with having the same velocity as the bulk motion of the 
cluster itself.
\item [{(2)}:] We derive the space velocity of the Hyades cluster by combining 
the radial velocities of the individual cluster members, their Hipparcos 
proper motions and their photometric distances (up to an initially
unknown global scale factor) derived from an isochrone of the Hyades main 
sequence.
All these independent data are elegantly combined in the statistical
 parallax method to derive a maximum likelihood estimate of the 
distance scale and the first two moments of the velocity distribution of the 
Hyades -- the bulk velocity and the velocity dispersion tensor
(\cite{hawley86}; \cite{strugnell86}; \cite{popowski98}).
This procedure generalizes the classical moving cluster methods.
\item [{(3)}:] We rescale the distance scale parameter of the statistical 
parallax solution from (2) to force agreement with the Hipparcos trigonometric 
parallaxes of the member stars in (1). This yields the best estimate of the
bulk velocity of the cluster
{\it under the assumption that Hipparcos parallaxes do not suffer from 
systematic errors}.
\item [{(4)}:] We adopt this common cluster velocity to predict the 
parallaxes of three Hyades spectroscopic binary systems whose orbital 
solutions are known accurately from previous work.
\item [{(5)}:] We compare the parallaxes predicted in this manner with the 
parallaxes from the orbital  solutions of the three binaries to check if 
there are any  systematic errors in the Hipparcos astrometry towards the
 Hyades cluster.
\end{description}

The outline of this paper is as follows.
We describe the various assumptions underlying our method in \S2.
We explain the connection between the classical moving cluster methods
and the statistical parallax method in \S3.
In \S4, we describe our selection of Hyades cluster members and use 
the statistical parallax algorithm to determine the common space velocity
and the distance to the centroid of the cluster.
We outline the procedure for estimating the parallaxes to individual
cluster members using the common cluster motion and proper motions from the 
\h catalog in \S5.
We compare the parallaxes predicted by this method with the parallaxes 
from the orbital solutions of the 3 binary systems in \S6.
We summarize our  conclusions in \S7 and describe the future potential of 
this technique.
As an aside, we note that we will drop the usual conversion factor 
$A_{v} = 4.74047 {\rm \ km}\, {\rm yr}\, {\rm s}^{-1}$ 
from all our equations for the sake of clarity, leaving it to the reader
to include it in the appropriate equations.

\section{ASSUMPTIONS}

We now describe the assumptions underlying the technique described
in this paper.
These include assumptions about both the kinematical structure of the 
cluster and the quality of the observational data.
These assumptions are:
\begin{description}
\item [{(1)}:] The velocity distribution of the Hyades cluster members
can be adequately described by an average bulk velocity and an isotropic
intrinsic velocity dispersion tensor.
In particular, we assume that the velocity structure of the cluster is
not significantly affected by shear, rotation or expansion.
From the Hyades age of $T \sim 625$ Myr (\cite{p98}), its velocity dispersion,
$\sigma\sim 330$ ${\rm m}\, {\rm s}^{-1}$, and its characteristic 
size $a\sim 5$ pc, we know that the cluster has survived for 
$\sigma T/a\sim 40$ crossing times. 
Hence, it is gravitationally bound and it is likely that all non-conserved 
modes of cluster motion such as radial pulsations (expansion and contraction),
shear etc., that may have been originally present, have by now been 
completely damped out.  
The one mode that could have survived is rotation because angular momentum 
is conserved, although Weyman (1967) finds that the rotation about
three mutually perpendicular axes is consistent with zero to within
$0.05$ ${\rm km}\, {\rm s}^{-1} {\rm pc}^{-1}$.
\item [{(2)}:] We assume a plausible shape for the isochrone
of the Hyades main-sequence.
However, this turns out to be not a very restrictive assumption since
we fit for the amplitude and the slope of the main-sequence using 
the photometric data of the Hyades members itself.
In principle, the shape of the isochrone could affect our selection of the 
Hyades members, although in practice, we find that the cluster membership 
remains unchanged for two different plausible isochrones.
\item [{(3)}:] We adopt the radial velocities collated from different
sources in the literature as the astrometric radial velocities of the 
stars themselves, even though these might include contributions from 
non-astrometric sources such as convective and gravitational line shifts,
atmospheric pulsations etc. (\cite{dravins86}; \cite{nadeau88}).
However, most of the radial velocities used by us come from the work of
Griffin et al. (1988), who calibrated the zero-point of their radial 
velocities using the radial velocities of asteroids and also applied
a magnitude dependent correction to remove the contributions from
convection in the stellar atmospheres (Gunn et al. 1988, hereafter G88).
In particular, $33$ stars of our $43$ Hyades members have their radial
velocities measured by Griffin et al. (1988).
\item [{(4)}:] We assume that the parallaxes of the three Hyades 
spectroscopic binary systems 51 Tauri, 70 Tauri and $\theta^{2}$ Tauri 
derived by Torres, Stefanik \& Latham (1997a, 1997b, 1997c, hereafter
T97a, T97b and T97c) are free from any systematic errors and that 
their quoted random errors are realistic.
\end{description}

\section{STATISTICAL PARALLAX AS A GENERALIZED MOVING CLUSTER METHOD}

In this section, we describe how the statistical parallax method is a
generalized form of the classical moving cluster methods and provides
a more precise distance to the Hyades cluster.
First, we present the various equations describing the geometry of the 
cluster motion and show how the statistical parallax method subsumes
the moving cluster methods.
We then use some simple assumptions about the Hyades cluster to 
present a quantitative estimate of how much  extra information about the
distance is present in the statistical parallax formalism compared to the
moving cluster methods.
However, while we present here a full statistical parallax solution
(including the distance scale), the only real use that we make of this 
solution in the present paper is in selecting the cluster members.
All the distance scale information in this solution is ``over-ridden'' by \h
trigonometric parallaxes before comparing with the binary orbital parallaxes 
so that the latter can directly test the former.
To make proper use of the statistical parallax Hyades distance measurement
would require a careful examination of systematic errors and we defer
this investigation to another paper.
We note that once we select the members (and adopt Hipparcos parallaxes),
the cluster space velocity does not depend on the details of the 
statistical parallax method at all.

Consider a cluster at a distance $d$ whose  bulk velocity is ${\bf V}$.
If the radial velocity at an appropriately defined cluster center is $V_{r}$
and the transverse velocity of the cluster in the plane of the sky is
${\bf V}_{T}$, we have,
\be
\mbox{\boldmath $\mu$} = \frac{{\bf V}_{T}}{d},
\label{eqn:mudef}
\ee
\be
{\bf V}_{T} = {\bf V} - V_{r}{\bf \hat r},
\label{eqn:vtdef}
\ee
where \mbox{\boldmath $\mu$} is the proper motion vector of the cluster center 
in the plane of the sky.
The difference between the transverse-velocity and the proper-motion vectors
($\delta {\bf V}_T {\rm \ and\ } \delta \mbox{\boldmath $\mu$}$)  of the cluster 
center and those of the individual cluster members are then given by
\be
\delta{\bf V}_{T} = - V_{r}\mbox{\boldmath $\theta$},
\label{eqn:delvtdef}
\ee
\be
\delta \mbox{\boldmath $\mu$} = \left( \frac{\delta{\bf V}_{T}}{d} \right)- \left( \frac{\delta d}{d} \right) \mbox{\boldmath $\mu$}
= -\left( \frac{V_{r}}{d} \right) \mbox{\boldmath $\theta$} - \left( \frac{\delta d}{d} \right)\mbox{\boldmath $\mu$},
\label{eqn:delmudef}
\ee
where $\mbox{\boldmath $\theta$}$ is the angular separation vector 
between the cluster center and the cluster member star in the plane of the sky,
and we have assumed that $\vert \mbox{\boldmath $\theta$} \vert \ll 1$ 
(the small angle approximation), and $(\delta d/d) \ll 1$.
This vector can be split into two components $\mu_{\parallel}$ and 
$\mu_{\bot}$ along the directions that are parallel and perpendicular 
respectively, to the  proper motion vector ($\mbox{\boldmath $\mu$}$) of the cluster in 
the plane of the sky.
Equation~(\ref{eqn:delmudef}) can then be written in terms of these 
components as
\be
\delta\mu_{\bot}  = -\left( \frac{V_{r}}{d} \right) \theta_{\bot},
\label{eqn:delmuperpdef}
\ee
\be
\delta\mu_{\parallel}  = -\left( \frac{V_{r}}{d} \right)\theta_{\parallel} -
\left( \frac{\delta d}{d} \right){\mu_{\parallel}},
\label{eqn:delmuparldef}
\ee
since ${\mu_{\bot}} = 0$, by definition.
Further, we also have 
\be
\delta V_{r} = (\mbox{\boldmath $\theta$} \cdot \mbox{\boldmath $\mu$})d = \theta_{\parallel} \mu_{\parallel}d =  \theta_{\parallel} V_{T},
\label{eqn:delvrdef}
\ee
where $\delta V_{r}$ is the difference between the radial velocities
 of the cluster member star and the cluster center, and 
$V_{T} = \vert {\bf V}_{T} \vert $.

It is clear from the above equations that there are three independent
measures of the distance to the cluster from
equations~(\ref{eqn:delmuperpdef}), (\ref{eqn:delmuparldef}), and 
(\ref{eqn:delvrdef}).
In the classical moving cluster method, the proper motions of the 
individual cluster members are used to derive a convergent point on the sky.
This information is combined with an average radial velocity of the 
cluster center to derive its  distance using equation~(\ref{eqn:delmuperpdef})
(\cite{boss08}; \cite{hanson75}; \cite{schwan91}).
Alternatively, if there are reliable radial velocities of the cluster members,
equation~(\ref{eqn:delvrdef}) can be used to estimate the cluster distance
by making use of the average proper motion of the cluster center 
(\cite{detweiler84}; G88).
So far, equation~(\ref{eqn:delmuparldef}) has been used only in a restricted
sense, in which the first term is used to derive a distance estimate,
while the second term is simply neglected assuming that 
the spread in the distance of individual stars ($\delta d$) 
adds to the uncertainty in this estimated distance (\cite{upton70}).
However, in this paper, we will use the photometric distance modulus to 
each star to estimate the quantity $(\delta d/d)$, and hence reduce the
uncertainty arising from the non-zero depth of the Hyades cluster.

All three independent estimates of the cluster distance are 
naturally combined in the statistical parallax method.
The resultant distance is then the weighted average of the individual 
distances from the three equations.
Since these distance estimates are independent of each other, their variances 
add harmonically. 
The weight from each of these estimates is given by 
$W_{i} = N(d_{i}/\sigma_{i})^{2}$ where $d_{i}$ and $\sigma_{i}, (i = 1,2,3)$
are the distances and the errors in the distances from each of the three
equations and $N$ is the total number of members that are
used to estimate the cluster distance.
These weights are approximately given by
\be
W_{1}  = N\left< \frac{(\theta_{\bot}V_{r})^{2}}{(d\sigma_{\mu})^{2} +\sigma^{2}} \right>,
\label{eqn:w1def}
\ee
\be
W_{2}  = N\left< \frac{(\theta_{\parallel}V_{r})^{2}}{(d\sigma_{\mu})^{2} +\sigma^{2} +(\sigma_{d}\mu)^{2}} \right>,
\label{eqn:w2def}
\ee
\be
W_{3}  = N\left< \frac{(\theta_{\parallel}V_{T})^{2}}{\sigma_{r}^{2} +\sigma^{2}} \right>,
\label{eqn:w3def}
\ee
where $\sigma_{r}$ and $\sigma_{\mu}$ are the errors in the radial velocities
and the proper motion respectively, $\sigma_{d}$ is the uncertainty in the 
relative distance to individual cluster members,
 and $\sigma$ is the velocity dispersion of the cluster.
The weight $W_{1}$ corresponds to the classical convergent-point 
moving cluster method using individual proper motions 
[eq. (\ref{eqn:delmuperpdef})], while $W_{2}$ corresponds to the extension
of this method using photometry to estimate the relative distances between
the cluster members [eq. (\ref{eqn:delmuparldef})].
The weight $W_{3}$ corresponds to the radial-velocity gradient method 
described by equation~(\ref{eqn:delvrdef})

For the purpose of illustration, we assume that for the Hyades cluster,
$\sigma = 0.3$ ${\rm km}\, {\rm s}^{-1}$,
$d\sigma_{\mu} = 0.3$ ${\rm km}\, {\rm s}^{-1}$, $\sigma_{r} = 0.2$ ${\rm km}\, {\rm s}^{-1}$,
 $\sigma_{d} \mu = 0.6$ ${\rm km}\, {\rm s}^{-1}$, 
$\left< \theta_{\parallel}^{2} \right>  = 
\left< \theta_{\bot}^{2} \right>  \equiv
\left< \theta ^{2}\right>,
$ and $V_{r} = (5/3)V_{T} = 40$ ${\rm km}\, {\rm s}^{-1}$.
This leads to $W_{1} : W_{2} : W_{3} = 1 : 0.33 : 0.5$,
showing that there is significant information about the 
distance in the two terms of equation~(\ref{eqn:delmuparldef}).
Here, we have assumed that $\sigma_{d} = 1.5 $ pc, arising from an error 
of $0.06$ mag in the photometric distance modulus of individual stars.
Hence, the distance estimate using the statistical parallax method is more
precise and accurate compared to that derived using the classical moving 
cluster methods alone.
We note that in the absence of observational errors, the fractional
accuracy in the cluster distance from the statistical parallax method
using $N$ cluster member stars is given by
\be
\frac{\Delta d}{d} = \sigma \left[\left( 2 V_{r}^{2} + V_{T}^{2} \right)N \left< \theta^{2} \right> \right]^{-1/2}.
\label{eqn:dfracerr}
\ee

\section{MEMBERSHIP AND COMMON CLUSTER MOTION}

We now find the bulk velocity of the Hyades cluster using the Hipparcos
proper motions of the cluster members.
However, a non-trivial problem here is the identification of the stars
belonging to the cluster itself.
We present our criteria for selecting the Hyades cluster members in 
\S4.1 and describe our method of deriving its space velocity in \S4.2.
In \S4.3, we compare our estimates of the cluster space velocity and
distance modulus with previous determinations of these quantities 
in the literature.

\subsection{\it Cluster Membership}

We select a preliminary set of 75 Hyades cluster candidates from the list of
282 Hyades candidates in Table 2 of Perryman et al. (1998, hereafter P98).
Our selection criteria for choosing these candidates are listed below
in the order in which they are enforced.
\begin{description}
\item [{(1)}:] We reject all the stars that are either known or suspected
 to be binaries from earlier work.
These are the stars with any alphabetical entry in at least one of the 
columns s, t or u of Table 2 of P98.
This criterion rejects a total of $130$ stars.
\item [{(2)}:] All the candidate stars should have radial velocity 
measurements.
We reject $14$ stars under this criterion.
We correct the raw radial velocity measurements of those stars measured 
by Griffin et al. (1988) (stars with an entry 1 in column r of Table 2 of P98) 
using the procedure described by equation~(12) of G88, but accounting 
for a sign error (\cite{p98}; R.Griffin 1998, private communication).
Of the $43$ cluster members used in this paper, we find that $33$ stars
have radial velocities measured by Griffin et al. (1988).
\item [{(3)}:] We reject any star that is flagged as a variable
in the Hipparcos catalog.
This condition eliminates one star, with Hipparcos ID HIP 17962.
\item [{(4)}:] We select only those stars that have ground based
photometric measurements of both $V_{J}$ and $(B-V)_{J}$.
We use the mean values of these quantities for the candidate 
stars from the GCPD photometric database of 
Mermilliod, Mermilliod \& Hauck (1997),
or from the Hipparcos catalog itself if the former are not available.
\item [{(5)}:] The candidate stars should have colors in the 
range $0.4 < (B-V)_{J} < 1.0$.
\end{description}
The last two photometric criteria (4) and (5) together eliminate
another $62$ stars, leaving us with $75$ single star Hyades candidates 
in the color range $0.4 < (B-V)_{J} < 1.0$, all of which have ground 
based photometric data, radial velocity data and Hipparcos astrometric data.
	
We derive the photometric distance modulus $(m-M)$ to each of the 75 candidate
stars in the color range $0.4 < (B-V)_{J} < 1.0$ by finding the difference
 between the apparent magnitude of the star and the absolute magnitude for its
color predicted by the isochrones of the Hyades main sequence.
For the adopted range in $(B-V)_{J}$ color, the isochrones are reliable
indicators of the distance modulus up to a possible global offset.
Since the Hyades isochrones have not previously been determined to high
precision, we apply our selection criteria using two distinct isochrones
which, as we show below, span the range of the true isochrone.
First, we use the isochrones  adopted by PSSKH98 and we 
refer the reader to that paper for further details about the construction
of the isochrones.
We assume a metallicity of $\left[{\rm Fe/H} \right] = +0.14$ and an
 age of $625 $ Myr for the Hyades (\cite{p98}).
We use the Yale color calibration (\cite{green88}) to transform the 
isochrones from the luminosity-temperature plane to the color-magnitude
plane.
Second, we  use the color calibration proposed by 
Alonso, Arribas, \& Martinez-Roger (1996) which predicts a different shape 
for the isochrone.
These two isochrones have different zero points and color dependence
with the result that if the isochrones are forced to 
coincide at $(B-V)_{J} = 0.4$, they differ by about 0.3 mag at 
$(B-V)_{J} = 1.0$.
We assume that the true isochrone is in the general range of these two
fiducial isochrones and parametrize it by the function
\be
M_{V}(B-V) = M_{V,{\rm Yale}}(B-V) + \Delta(m-M) + \alpha [ (B-V)_{J} - 0.7]
\label{eqn:isodef}
\ee
where $\Delta(m-M)$ and $\alpha$ are parameters to be determined.
These allow for both an offset in the zero point and a different slope for the
color-magnitude relation.

For each pair of values of $\Delta(m-M)$ and $\alpha$, we determine the 
space velocities (${\bf V}_{i}$) of all these candidates using their
photometric distance modulus, their radial velocities and their proper motions
from the Hipparcos catalog.
We derive a best-fit mean velocity (${\bf \bar V}$) of their centroid and 
reject the stars that are gross outliers from this mean cluster motion.
We compute the quantity $\chi^{2}$ defined as,
\be
\chi^{2}  = \sum_{i=1}^{N}\left( {\bf V}_{i} - {\bf \bar V} \right)^{T} {\bf C}_{i}^{-1} \left( {\bf V}_{i} - {\bf \bar V} \right),
\label{eqn:chi2}
\ee
where the summation is over all the N stars that remain at each iteration.
The covariance matrix ${\bf C}_{i}$ for star $i$ includes contributions from
the error  in the photometric distance modulus, 
from the error in the radial velocity, from the error in the proper motion, 
and from the velocity dispersion of the cluster.
The cluster velocity dispersion is between $0.2$ to $0.4$ ${\rm km}\, {\rm s}^{-1}$ 
(\cite{gunn88}; P98; D97) for plausible values 
of the cluster mass of about 300$M_{\odot}$ to 450$M_{\odot}$  and half-mass 
radius of the cluster of about 4 to 5 pc (\cite{pels75}; \cite{gunn88}, P98).
In the first iteration of our membership selection, we assume a value of 
$0.4$ ${\rm km}\, {\rm s}^{-1}$.
We estimate the error in the distance modulus as arising solely due to the
error in the $(B-V)_{J}$ color and assume an average  slope of 6 for the 
isochrone to translate this to an error in $(m-M)$.

We iterate the procedure described above until there are no strong outliers 
and the velocities of all the remaining stars are consistent with a common 
cluster motion.
At each iteration, we include only those stars with photometric 
distance modulus $(m-M) \le 4.5 + \Delta(m-M)$, thus excluding stars
that are located more than $30$ pc beyond the centroid of the cluster.
This procedure eliminates $13$ stars from our original list of $75$
candidates, of which $12$ are classified as non-members by P98.
The remaining one star (with Hipparcos ID HIP 20205) is a giant 
(spectral class G8III according to the Hipparcos catalog) and we do not 
include it in our distance determination as its distance modulus cannot
be estimated using an isochrone of the Hyades main sequence.
We reject any star whose individual contribution to $\chi^{2}$ 
is greater than $(3\chi^{2}/N)$ as an outlier.
We derive our best estimates of $\alpha$ and $\Delta(m-M)$ from the
remaining stars and use these values to estimate the cluster distance, 
its bulk velocity, and its velocity dispersion ($\sigma$) using the statistical
parallax algorithm.
These two steps are described in detail in \S4.2.
The statistical parallax method also yields the extra error in the 
photometric distance modulus ($\epsilon$) which should be added in quadrature
to the quoted errors so that the velocity dispersion tensor is isotropic.
We repeat the cluster membership using these new values of
$\alpha$, $\Delta(m-M)$, $\sigma$ and $\epsilon$.
We iterate the entire procedure until the cluster membership converges.

We present the results of our iterative procedure in Table \ref{table:tab1} in which we
list the number of stars used at each iteration ($N_{\rm star}$), 
the total $\chi^{2}$ at the end of each iteration, the threshold value of 
the individual $\chi^{2}$ of every star below which we accept the 
candidates as members $( = 3\chi^{2}/N_{\rm star})$, the number of stars that 
are rejected as non-members at each iteration ($N_{\rm reject}$,) and the 
number of rejected stars that are classified as members by P98 $(N_{\rm P98})$.
Of the remaining $62$ Hyades candidates, $11$ stars are classified as 
non-members by  both P98 and by our algorithm.
There are an additional $8$ stars (with Hipparcos IDS HIP 20187, 21788, 16908,
21112, 21267, 20491, 21066, 13806) that are classified as members by P98, 
but which we reject as non-members because their individual contributions to 
the total $\chi^{2}$ are $235.2, 230.5, 19.4, 18.0, 16.4, 14.9, 14.9$, and
$11.9$ respectively, while the threshold value of $(3\chi^{2}/N) = 9.9$ at the 
end of the last iteration.
We note that even if we increase the threshold to $(4\chi^{2}/N)$, only one
additional star (HIP 13806) will be classified as a member.
On the other hand, the $5$ highest individual contributions to $\chi^{2}$ of 
our member stars are $9.3, 9.1, 7.0, 6.7$ and $6.7$ (corresponding to the stars
HIP 19504, 20130, 22566, 20082 and 20349).

We select a total of $43$ stars as Hyades cluster members at the end of
the last iteration, all of which lie in a tight cluster around the 
mean cluster motion in velocity space.
The best fit values of the parameters at the last iteration are
$\alpha = 0.28 \pm 0.07$, $\Delta(m-M) = 0.10 \pm 0.03$ mag,
$\sigma = 320 \pm 39 {\rm \ m}\, {\rm s}^{-1}$, and $\epsilon = 0.042$ mag.
Our membership selection procedure is robust to any changes in the absolute 
calibration of the isochrones since the relative distances between the cluster 
candidates are unaffected by this.
However, it is sensitive to the shape of the isochrones, although in practice,
we find that the cluster membership is the same for the two fiducial 
isochrones, despite the fact that they span  a much larger range than is
allowed by our fits for $\Delta(m-M)$ and $\alpha$ (see \S4.2).

\begin{table}
\begin{center}
\caption{Cluster membership iterations}
\bigskip
\begin{tabular}{*{6}{c}}
\tableline\tableline
Iteration & $N_{\rm star}$ & $\chi^{2}$ & $(3\chi^{2}/N_{\rm star})$& $N_{\rm reject}$ & $N_{\rm P98}$ \\
\tableline
1 & 62 & 23282 & 1127 & 6 & 0 \\
2 & 56 & 3947 &  212 & 5 & 0 \\
3 & 51 & 682.3 &  40.14 & 2 & 2 \\
4 & 49 & 235.2 &  14.4 & 2 & 2 \\
5 & 46 & 182.2 &  11.9 & 2 & 2 \\
6 & 44 & 153.8 &  10.5 & 1 & 1 \\
7 & 43 & 142.5 &  9.9 & 0 & 0 \\
\tableline
\end{tabular}
\label{table:tab1}
\end{center}
\end{table}

\subsection{\it Distance and Space velocity of Hyades}

We determine the common space velocity of the cluster from the velocities
of all the cluster members selected by the procedure described in \S4.1.
We evaluate the $\chi^{2}$ [as defined in eq. (\ref{eqn:chi2})] in a dense 
grid of points in the space of the five parameters namely, $\Delta(m-M)$, 
$\alpha$ and the three components of the bulk velocity (in equatorial
coordinates) of the cluster $({\bf \bar V})$.
We fit this to a quadratic in the 5 parameters to find the 
best fit values and the covariance matrix of the parameters at the 
minimum of the $\chi^{2}$ surface.
We adopt this best-fit value of the bulk velocity as our initial guess
of the cluster space velocity in the statistical parallax method.
We estimate the components of the cluster space velocity in a coordinate 
system that is oriented such that one of the axes is along the radial 
direction of the centroid ($V_{r}$), another axis is along the direction
 perpendicular to the proper motion of the cluster in the plane of the sky 
($V_{\bot}$) and the third axis is parallel to the proper motion of the 
cluster in the plane of the sky ($V_{\parallel}$).
By definition, $V_{r}$ is the radial velocity of the cluster,
 $V_{\parallel}$ is its velocity in the plane of the sky, and 
$V_{\bot}$ is zero.

To compute the photometric distance to each star, we fix the 
slope-correction $(\alpha)$ and the zero point offset $[\Delta(m-M)]$ at 
their best-fit values derived as described above. 
In the modern version of the statistical parallax method as described by 
Popowski \& Gould (1998), one uses the maximum likelihood procedure to 
simultaneously solve for ten different parameters viz., the distance 
scaling factor relative to a fiducial distance scale ($\eta$), the three 
components of the bulk velocity of the cluster ($V_{r}, V_{\bot}$ and 
$V_{\parallel}$) and the six independent components of the second moments 
of its velocity distribution -- the three diagonal terms corresponding to
 the square of the velocity dispersion in the three  directions 
($\sigma_{r}^{2}, \sigma_{\bot}^{2}$ and $\sigma_{\parallel}^{2}$) and 
the three unique off-diagonal terms ($\sigma_{r\bot}^{2}, 
\sigma_{r\parallel}^{2}$ and $\sigma_{\perp \parallel}^{2}$).
We assume an isotropic velocity dispersion tensor of the Hyades with the 
result that the three independent off-diagonal terms are constrained to 
be zero.
For an assumed level of errors in the radial velocities, the proper 
motions, and the distance to individual stars, the statistical parallax method 
derives a maximum likelihood estimate of the cluster velocity dispersions in 
the three mutually perpendicular directions.
However, the errors in the distance to each star affect only the parallel
dispersion $\sigma_{\parallel}$ while the estimates of the velocity 
dispersions in the radial and the perpendicular directions ($\sigma_{r}$ and
$\sigma_{\bot}$ respectively) are independent of the distance errors.
Therefore, we begin by constraining the velocity dispersion in these two 
directions to have the same value.
The velocity dispersion in the parallel direction ($\sigma_{\parallel}$) now 
includes contributions from both the intrinsic velocity dispersion of the
cluster and the dispersion arising from a possibly wrong estimate of the 
distance errors.
Hence, we add an extra error ($\epsilon$) in quadrature to our 
original errors (as listed in the photometric sources) in the photometric distance modulus so that the velocity dispersion in the parallel direction 
($\sigma_{\parallel}$) becomes equal to the velocity dispersions in the 
other two directions.
We adopt this velocity dispersion as the velocity dispersion of the 
cluster.

At the end of the last iteration, the sample contains $43$ Hyades cluster 
members, the velocity dispersion is equal to 
$320 \pm 39$ ${\rm m}\, {\rm s}^{-1}$,
and we need to add an extra error of $\epsilon = 0.042$ mag (in quadrature) 
in the photometric distance modulus to enforce an isotropic velocity dispersion
tensor.
We adopt this value of the velocity dispersion and $\epsilon$ 
in the remainder of this paper.
The statistical parallax method finds the maximum likelihood solution of the
five independent parameters namely,
 ${\bf p} = (\eta, V_{r}, V_{\bot}, V_{\parallel},\sigma_{r}^{2})$ subject 
to the constraint of an isotropic velocity dispersion tensor.
This solution, using only the relative photometric distances to individual
Hyades cluster stars is given by 
${\bf p}({\rm stat}) = [ 1.0027 \pm 0.0143, 39.50 \pm 0.06 {\rm \ km}\, {\rm s}^{-1},
0.00 \pm 0.07 {\rm \ km}\, {\rm s}^{-1}, 23.82 \pm 0.34 {\rm \ km}\, {\rm s}^{-1},
0.1034 \pm 0.0251 \ ({\rm km}\, {\rm s}^{-1})^{2}]$, 
where the velocity
dispersion is, as expected, $\sigma_{r} = 320 \pm 39$ ${\rm m}\, {\rm s}^{-1}$.

The statistical parallax measurement of the distance scale $(\eta_{\rm phot})$
is of interest in its own right.
However, a proper interpretation of this measurement requires a thorough
investigation of the systematic errors which is beyond the scope of
this paper.
Here, our primary goal is to test the Hipparcos {\it trigonometric parallax}
distance scale toward the Hyades.
To this end, we will therefore {\it renormalize} $\eta$ to force agreement
(on average) between the distances to cluster members as determined in the 
renormalized solution and from \h parallaxes.
This will have the effect of rescaling $V_{\parallel}$ by the same factor,
but will have no effect on our estimate of $V_{r}$ or $V_{\bot}$.
The resulting space velocity of the cluster $(V_{r}, V_{\bot}, V_{\parallel})$
can then be used to measure the distances to individual Hyades binary 
systems from their proper motions and thus test the \h trigonometric-parallax 
Hyades distance scale against the orbital parallaxes of these
binaries.

We begin by computing for each cluster member, the quantity
\be
\eta_{i} = \frac{\pi_{{\rm phot},i}}{\pi_{{\rm Hip},i}}\left(1+x_{i}^{2}\right),
\label{eqn:etaidef}
\ee
where 
\be
x_{i} = \frac{\sigma_{\pi, {\rm Hip}, i}}{\pi_{{\rm phot},i}}.
\label{eqn:xidef}
\ee
Since the errors in the \h parallaxes are approximately Gaussian distributed
(\cite{arenou95}),
the quadratic correction term $x_{i}^{2}$ is required to ensure that the 
two sides of equation~(\ref{eqn:etaidef}) have the same mean value
(\cite{lk73}; \cite{smith96}).
The error in $\eta_{i}$ is given by
\be
\sigma_{{\eta},i} = \left( \frac{\sigma_{\pi, {\rm Hip}}}{\pi_{{\rm phot}}} \right)_{i} \eta_{i}.
\label{eqn:etaierror}
\ee
We find that the mean value of $\eta$ for all the cluster members is 
given by $\eta_{\rm Hip} = 1.0178 \pm 0.0081$.
This is consistent with the value of $\eta_{\rm phot} = 1.0027 \pm 0.0181$
derived using the \h proper motions alone.
The difference between these two distance scaling factors is given by
\be
\eta_{\rm phot} - \eta_{\rm Hip} = - 0.0151 \pm 0.0164,
\label{eqn:etaphothipdiff}
\ee
and hence is consistent with zero, thus providing a semi-independent
check of the self-consistency of \h astrometry.
Although P98 reached a similar conclusion, our result is based on 
improved parallax estimates, as the statistical parallax solution
also includes the photometric distance information to individual stars.
This consistency between the scaling factors would permit us, if we
desired, to combine the two determinations to find a single best estimate
of $\eta$.
However, our aim here is to test the \h parallaxes alone.
Therefore, we simply constrain the overall solution to have 
$\eta = \eta_{\rm Hip} = 1.0178 \pm 0.0081$.
We then find that the space velocity of the cluster is 
$(V_{r}, V_{\bot}, V_{\parallel}) = (39.51 \pm 0.06 {\rm \ km}\, {\rm s}^{-1},
0.00 \pm 0.07 {\rm \ km}\, {\rm s}^{-1}, 24.17 \pm 0.22{\rm \ km}\, {\rm s}^{-1})$ 
and the matrix  of correlation coefficients is
\be
\left( 
\begin{array}{rrrrr}
1.0000 & -0.0045 & 0.0161 \\
-0.0045 & 1.0000 & -0.1392 \\
0.0161 & -0.1392 & 1.0000 \\
\end{array}
\right).
\label{eqn:pcovar}
\ee
The two primary effects of using the \h parallaxes to fix the 
distance scale to the Hyades members are:
\begin{description}
\item [{(1)}:] It increases the value of $V_{\parallel}$ by $1.5 \%$
i.e., almost exactly equal to the increase in the value of $\eta$ itself.
\item [{(2)}:] It significantly reduces the error in $V_{\parallel}$.
\end{description}
However, neither $V_{r}$, nor $V_{\bot}$, nor their errors are significantly
affected.
The fact that the maximum likelihood value of $\eta$ is greater than one
means that the \h parallax distance scale to the Hyades is larger than our 
fiducial scale by a factor $(\eta-1) = 1.78\%$.
The space velocity of the cluster in equatorial coordinates is
$(V_{x}, V_{y}, V_{z}) = (-5.70 \pm 0.20, 45.62 \pm 0.11, 5.65 \pm 0.08)$ ${\rm \ km}\, {\rm s}^{-1}$ and the matrix of correlation coefficients is
\be
\left( 
\begin{array}{rrr}
 1.0000 & -0.8134 & 0.5042 \\
-0.8134 & 1.0000 & -0.4964 \\
0.5042 & -0.4964 & 1.0000 \\
\end{array}
\right).
\label{eqn:vcovar}
\ee
We will use this estimate of the bulk velocity of the cluster 
in the remainder of the paper to predict the parallaxes
from the Hipparcos proper motions of individual stars.
In Galactic coordinates, this velocity is given by
$(U, V, W) =  (-42.27 \pm 0.08, -18.89 \pm 0.18, -1.51 \pm 0.14)$ ${\rm \ km}\, {\rm s}^{-1}$.

We show the velocities of the Hyades cluster candidates in Figure 1.
The first three panels (a)-(c) show the velocities computed using the 
photometric distance modulus to each star (normalized to the \h 
trigonometric parallax distance scale) and the Hipparcos proper motions,
while panel (d) shows the velocities computed using the Hipparcos parallaxes
and proper motions.
The crosses show the velocity components of the 43 cluster members, while 
the open circles represent the velocity components of the $8$ stars that are 
classified as members by P98, but rejected as non-members by our algorithm.
The smaller scatter in velocities of the members in panel (c) compared to
 that in panel (d) shows that the photometric distance moduli lead to a much 
tighter core in velocity space, and hence a cleaner separation between 
the members and the non-members compared to using distances inferred
from Hipparcos parallaxes.
The solid circle in all the panels shows the bulk velocity of the cluster.
For this space velocity of the cluster, the total $\chi^{2}$ is 143 for 43
 stars (corresponding to 126 degrees of freedom) demonstrating that our 
estimates of the errors for the various quantities and of the cluster
 velocity dispersion are reasonable.
The centroid is at a distance of 
$\vert {\bf \bar r} \vert = 46.61 \pm 0.38$ pc (corresponding to 
a distance modulus of $3.34 \pm 0.02$), and its
equatorial coordinates are $\alpha = 04^{h}26^{m}32^{s}, 
\delta = 17^{\circ}13.\hskip-2pt'3 $ (2000).
This is also the direction of the radial velocity of the cluster center,
i.e, the direction of $V_{r}$.
The motion of the cluster in the plane of the sky is towards the direction
$105^{\circ}11.\hskip-2pt'0$ East of North.

\begin{figure}
\centerline{
\epsfxsize=\hsize
\epsfbox[18 144 592 718]{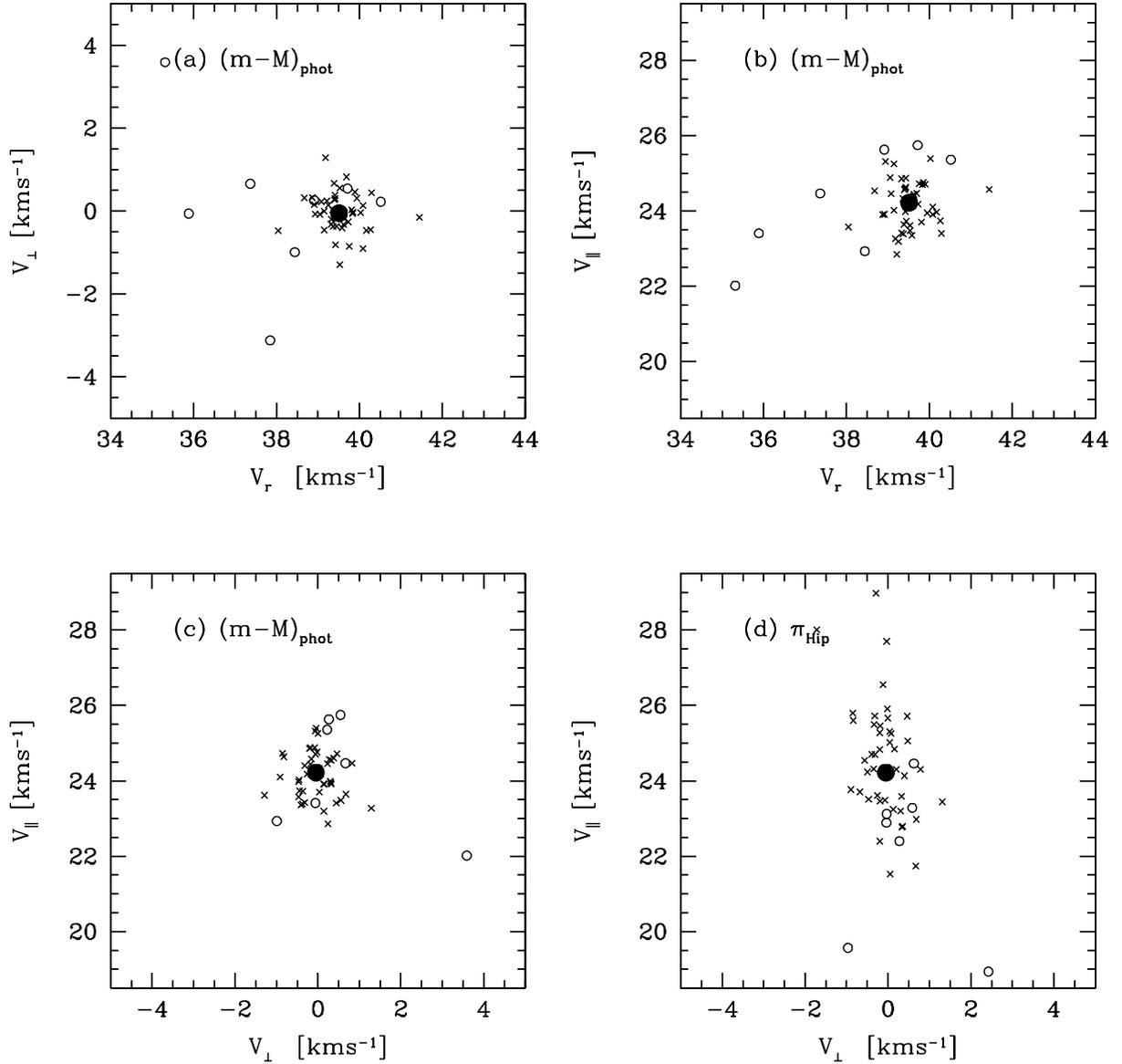}
}
\caption{Velocities of the Hyades cluster candidates from the \h proper 
motions.
The velocities in panels (a)-(c) are computed using the photometric distance
to each star (normalized to the \h trigonometric parallax distance scale), 
while those in panel (d) are computed using the distance inferred directly 
from the Hipparcos parallax.
(a) $V_{r}$ and $V_{\perp}$
(b) $V_{r}$ and $V_{\parallel}$ 
(c) $V_{\perp}$ and $V_{\parallel}$ and
(d) $V_{\perp}$ and $V_{\parallel}$.
The component $V_{r}$ is the velocity component in the radial 
direction of the centroid, while $V_{\bot}$ and $V_{\parallel}$ represent
the velocity components perpendicular and parallel to the direction of the 
proper motion of the cluster in the plane of the sky.
The crosses are the velocity components of the 43 cluster members while 
the open circles are those of the $8$ stars that are  classified as members
by P98, but rejected as non-members by our algorithm.
The solid circle in each panel shows the mean motion of the cluster.
}
\end{figure}

\subsection{\it Comparison with previous estimates}

We compare our estimates of the distance modulus to Hyades centroid,
its bulk space velocity, and its velocity dispersion,
with previous determinations of these quantities in the literature.

Our distance modulus to Hyades of $(m-M) = 3.34 \pm 0.02$ mag agrees very
well with the value of $3.33 \pm 0.01$ mag obtained by P98 using the stars
located within $10$ pc of the cluster center, and the value of $3.34 \pm 0.04$
mag determined by PSSKH98 using the main-sequence fitting technique.
We would like to emphasize here that while the P98  value refers
to the distance modulus of the center of mass of the Hyades, our estimate 
refers to the centroid of the $43$ Hyades members selected in \S4.1.
However, the estimate of the bulk velocity does not depend on the choice 
of the subsample of the cluster and our results can therefore be directly 
compared to those of P98 and D97.

The space velocity of the Hyades cluster has been recently determined
by P98 and by D97 using the Hipparcos astrometric data of the Hyades members.
We compare these velocity estimates with our estimate of the bulk velocity
of the cluster in the coordinate system centered on the centroid of Hyades,
viz., $(V_{r}, V_{\bot}, V_{\parallel}) = (39.51 \pm 0.06 {\rm \ km}\, {\rm s}^{-1},
0.00 \pm 0.07 {\rm \ km}\, {\rm s}^{-1}, 24.17 \pm 0.22 {\rm \ km}\, {\rm s}^{-1})$.
In this coordinate frame, the cluster velocity determined by P98 is
$(38.82, 0.03, 24.55) {\rm \ km}\, {\rm s}^{-1}$, while that determined by 
D97 is $(39.60, 0.00, 24.65) {\rm \ km}\, {\rm s}^{-1}$.
The errors in these velocity components are unlikely to exceed
$0.2 {\rm \ km}\, {\rm s}^{-1}$.
Our estimate of the radial velocity of the Hyades centroid is significantly
larger than that of P98, but agrees very well with the value of D97.
However, the first two panels in Figure 1 show that the radial velocities 
of $5$ of the $8$ stars that are classified as members by P98, but rejected as 
non-members by  our algorithm are significantly 
(more than $1 {\rm \ km}\, {\rm s}^{-1}$) smaller than the mean
radial velocity of the cluster, and including them as members will 
systematically reduce the mean cluster radial velocity.
These $5$ stars are clearly well outside the tight cluster in the radial
velocity component $(V_{r}$), have high values of $\chi^{2}$, 
and are therefore unlikely to be Hyades members given that the 
cluster velocity dispersion is only $320 \pm 39$ ${\rm m}\, {\rm s}^{-1}$.
Although the radial velocities of the other $3$ stars 
(HIP 13806, 21066 and 21112) are consistent with them being Hyades members, 
they have systematically high values of the parallel  velocity component
($V_{\parallel}$) and  are therefore classified as non-members.
However, the star HIP 20350 whose radial velocity is about 
$41.44 {\rm \ km}\, {\rm s}^{-1}$ is accepted as a Hyades member because of 
the large error in its radial velocity 
($\sigma_{r} = 2.4 {\rm\ km}\, {\rm s}^{-1}$).
Hence, the difference between our estimate of the radial velocity and that
of P98 can be attributed to the differences in the selection of Hyades 
members.

All the estimates of the velocity component in the perpendicular direction
($V_{\bot}$) are consistent with each other, and are individually consistent
with the expected value of zero.
However, our estimate of the parallel velocity component 
$(V_{\parallel} = 24.17 \pm 0.22$) is inconsistent at the $1.7 \sigma$
level with the value derived by P98, and at the $2.2 \sigma$ level with that
derived by D97.
The cluster velocity dispersion of $320 \pm 39 {\rm \ m}\, {\rm s}^{-1}$
computed using our statistical parallax algorithm agrees well with the 
value of $300 {\rm \ m}\, {\rm s}^{-1}$ determined by P98 and acceptably with
the value of $250 \pm 40 {\rm \ m}\, {\rm s}^{-1}$ determined by D97, although
our value is computed excluding all the binary systems that dominate the 
mass distribution near the central regions of the cluster
(P98, \cite{pels75}; \cite{terlevich87}).

\section{PARALLAX FROM PROPER MOTION}

We adopt the cluster space velocity derived in the previous section
to predict the parallaxes
of the 43 member stars using their proper motions from the \h catalog.
The parallax of any cluster member that has the same space velocity as the 
cluster is given by
\be
\pi_{{\rm pm},i} = \frac{\left< ({\bf V_{t}})_{i} \vert {\bf C}_{i}^{-1} \vert \mbox{\boldmath $\mu$}_{i} \right> }{\left< ({\bf V_{t}})_{i} \vert {\bf C}_{i}^{-1} \vert ({\bf V_{t}})_{i} \right>}
\label{eqn:pipm}
\ee
where 
$({\bf V_{t}})_{i} = {\bf V_{c}} - (\hat {\bf r}_{i} \cdot {\bf V_{c}})\hat{ \bf r}_{i} $ 
is the transverse velocity of the cluster in the plane of the sky at the
position of the star $i$, $\mbox{\boldmath $\mu$}_{i}$ is its proper motion
 from the \h catalog and ${\bf C}_{i}$, the covariance matrix, is the sum of
the proper motion error tensor of star $i$ and the velocity dispersion
 tensor divided by the square of the distance.
We have employed Dirac notation,
\be
\left< X \vert {\mathcal{O}} \vert Z \right> = \sum_{i,j}X_{i}{\mathcal{O}}_{ij}Z_{j}.
\label{eqn:diracdef}
\ee
The error in $\pi_{{\rm pm},i}$ is equal to 
$\left< ({\bf V_{t}})_{i} \vert {\bf C}_{i}^{-1} \vert ({\bf V_{t}})_{i} \right>
^{1/2}$.

We show the difference between the photometric parallax ($\pi_{\rm {phot}}$) 
and the \p determined assuming a common space velocity for all the 
cluster members ($\pi_{\rm pm}$) in Figure 2.
We have scaled all the photometric distances by the factor
$\eta_{\rm Hip} = 1.0178$ so that the mean value of this difference
 should be equal to zero.
The horizontal error bars show the error in photometric parallax, while
the vertical error bars show the uncertainty in $\pi_{{\rm pm}}$ alone.
Thus, the total error in the quantity $(\pi_{\rm phot} - \pi_{\rm pm})$
is the sum in quadrature of the two error bars displayed in Figure 2.
We find that the mean offset between the two parallaxes is given by
\be
\left< \pi_{\rm phot} - \pi_{\rm pm} \right> = -0.009 \pm 0.099 {\rm\ mas},
\label{eqn:photpmset}
\ee
while the $\chi^{2}$ of this difference is $29.2$ for a total of $43$ stars.
This shows that there are no internal inconsistencies in our method of 
predicting the parallaxes from the proper motions.
\begin{figure}
\centerline{
\epsfxsize=\hsize
\epsfbox[18 144 592 718]{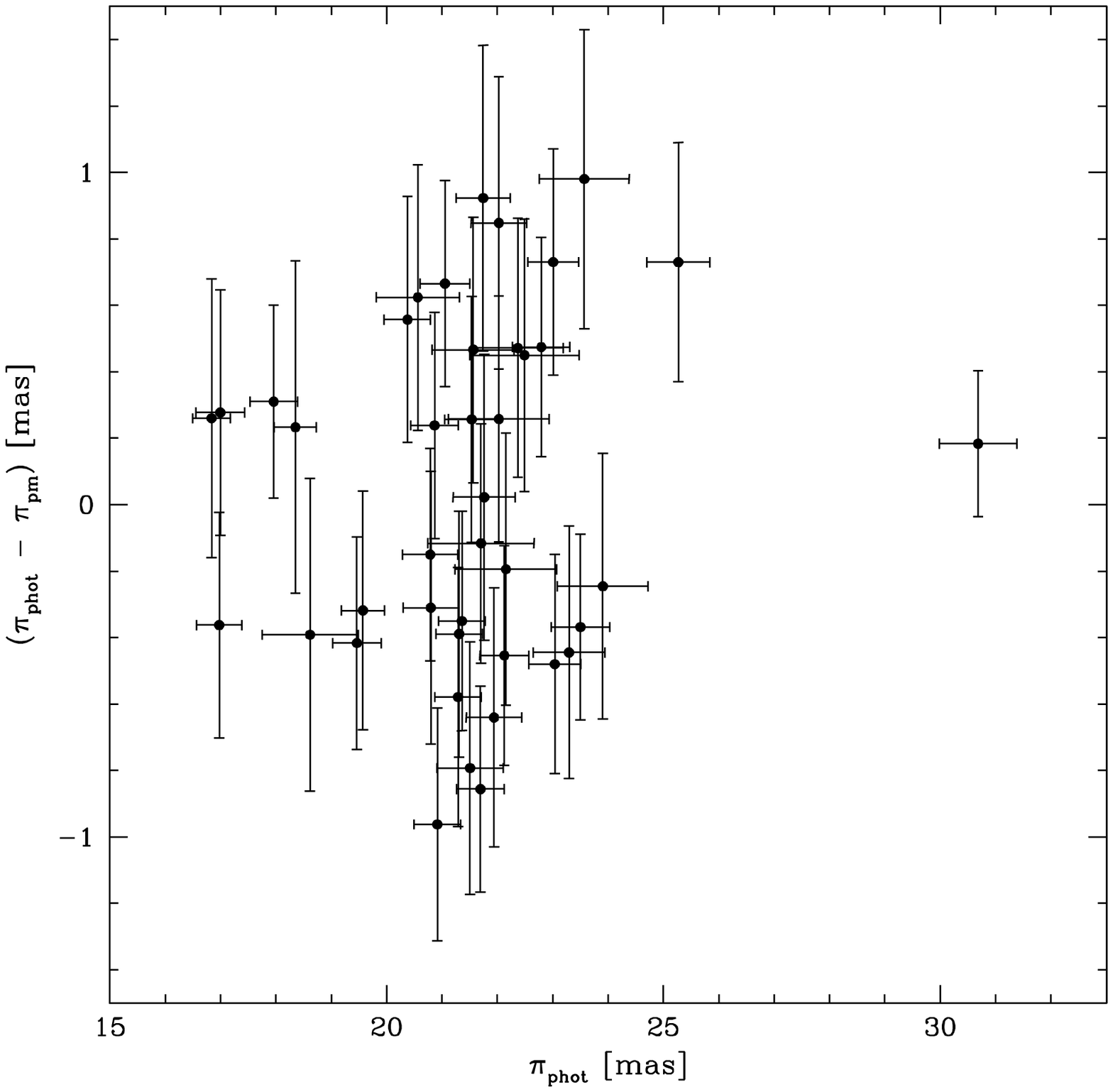}
}
\caption{Difference between the photometric parallax ($\pi_{\rm phot}$)
and the parallax predicted assuming a common space velocity for the cluster 
members $(\pi_{\rm pm})$.
All the photometric distances have been scaled by the quantity 
$\eta_{\rm Hip} = 1.0178$  so that the mean value of this difference
should be zero.
The horizontal error bars show the error in $\pi_{\rm phot}$, while
the vertical error bars show the uncertainty in $\pi_{\rm pm}$ alone, i.e, the
error in ($\pi_{\rm phot} - \pi_{\rm pm}$)  is the quadrature sum of the
two error bars.
}
\end{figure}
\section{PARALLAXES OF BINARY SYSTEMS}

We predict the parallaxes of the 3 binary systems 51 Tauri
 (HIP 20087), 70 Tauri (HIP 20661) and $\theta^{2}$ Tauri (HIP 20894) using 
the cluster space velocity $({\bf V_{c}})$ determined above and their
individual Hipparcos proper motions.
The full orbital solutions of these 3 spectroscopic binary systems have
been derived by T97a, T97b and T97c respectively.
The Hipparcos proper motions refer to the center of mass for HIP 20087, 
while they refer to the motion of the photocenter for the other two systems.
HIP 20894 whose semi-major axis is less than $0.\hskip-2pt''1$ is listed
as a variable single star in the Hipparcos catalog.
For HIP 20661 and HIP 20894, we compute the difference between the proper 
motion of the center of mass and the proper motion of the photocenter
using the spectroscopic-astrometric orbital solutions for these 2 binary
 systems provided by T97b and T97c respectively.
In Table 2, we list the proper motions of the center of mass of all the 
3 systems, their parallaxes from Hipparcos, their parallaxes from their
 proper motions, and their orbital parallaxes.
There are two important features in the errors of the different parallaxes 
in Table \ref{table:tab2}.
\begin{description}
\item [{(1)}:] The error in the parallaxes determined from the individual 
proper motions of the binary systems and the common space velocity of the 
cluster [$\sigma_{\pi}({\rm pm})$ in column 7 of Table 2] is almost a factor 
of three smaller than the error in the Hipparcos parallaxes.
\item [{(2)}:] The errors in the orbital parallaxes of these binary 
systems [$\sigma_{\pi}({\rm orb})$ in column 9 of Table 2]
are about twice as large as the errors in the proper-motion 
parallaxes.
\end{description}
\begin{table}
\begin{center}
\caption{Astrometry of the 3 spectroscopic binary systems with orbital parallaxes.}
\bigskip
\begin{tabular}{*{9}{c}}
\tableline\tableline
HIP ID & $\mu_{\alpha}{\rm cos(\delta)}$ & $\mu_{\delta}$ & $\pi_{\rm Hip}$ & $\sigma_
\pi({\rm Hip})$ & $\pi_{\rm pm}$ & $\sigma_\pi({\rm pm})$ & $\pi_{\rm orb}$ & $\sigma_
\pi({\rm orb})$ \\
& (mas yr$^{-1}$) & (mas yr$^{-1}$) & (mas) & (mas) & (mas) & (mas) & (mas) & (mas) \\
 \tableline
20087 & 96.42 & -33.92 & 18.25 & 0.82 & 18.45 & 0.30 & 17.92 & 0.58 \\ 
20661 & 104.97 & -26.67 & 21.47 & 0.97 & 21.35 & 0.36 &  21.44 & 0.67 \\
20894 & 108.80 & -26.35 & 21.89 & 0.83 & 22.49 & 0.37 &  21.22 & 0.76 \\ \tableline
\end{tabular}
\label{table:tab2}
\end{center}
\end{table}

The mean  difference (weighted by the inverse square errors) between the 
proper-motion parallaxes and the  orbital parallaxes for these three binary 
systems is,
\be
\left< \pi_{\rm pm} - \pi_{\rm orb} \right> = 0.52 \pm 0.43 {\rm \ mas}.
\label{eqn:pmorboffset}
\ee
The error in this difference is dominated by the error in the
 orbital parallaxes.
From Table 1, we find that the error in the individual proper-motion
 parallaxes are each of the order of 0.3 to 0.35 mas.
Hence, it appears possible in principle to detect any systematic errors in the 
Hipparcos parallaxes at the level of  
$\sigma_{\rm sys} = \left[\sum 1/\sigma_{\pi}^{2}(\rm pm)\right
]^{-1/2} = 0.2 $ mas
if the orbital parallax errors could be reduced below this level.
However, there is another source of error in determining the proper-motion 
parallaxes.
This arises from the error in the component of the cluster space velocity 
itself in the direction of its proper motion in the plane of the sky, i.e, in 
the component $V_{\parallel}$.
This error is equal to 
$\left< \pi_{\rm Hya} \right> (\sigma_{V_{\parallel}}/V_{\parallel}) = 0.19$ mas
where we have again assumed that the mean \p of the Hyades cluster is 21.5 mas.
Adding this error in quadrature to the errors determined above, we find that
$\left< \pi_{\rm pm} - \pi_{\rm orb} \right> = 0.52 \pm 0.47 {\rm\ mas}$.
The irreducible error in this method $\sigma_{\rm sys}$ is now 0.28 mas, 
still considerably smaller than the errors in the currently available 
orbital parallaxes.
We note that this analysis assumes that there are no systematic errors
in the orbital parallaxes of the binary systems.

P98 also concluded that the Hipparcos parallaxes of these binary systems 
and their parallaxes from their orbital solutions are consistent.
However, the accuracy of their comparison is limited by the errors in 
the Hipparcos parallaxes.
Thus, we find that the mean  difference (weighted by the inverse square 
errors) between the Hipparcos parallaxes and the  orbital parallaxes for 
these three binary systems is
\be
\left< \pi_{\rm Hip} - \pi_{\rm orb} \right> = 0.35 \pm 0.63 {\rm \ mas}.
\label{eqn:hiporboffset}
\ee
The error in this difference is dominated by the error in the 
Hipparcos parallaxes and is therefore irreducible in the future.
This shows that the technique proposed in this paper allows one 
to check the level of systematic errors in the \h parallaxes at more than 
twice the precision than is possible using the straightforward comparison 
of \h and orbital parallaxes.

\section{CONCLUSIONS}

Our main conclusions are as follows:
\begin{description}
\item [{(1)}:] When the distance scale to the Hyades cluster is fixed by
\h parallaxes, the bulk velocity in equatorial  coordinates is
$(V_{x}, V_{y}, V_{z}) = (-5.70 \pm 0.20, 45.62 \pm 0.11, 5.65 \pm 0.08)$ ${\rm \ km}\, {\rm s}^{-1}$, its velocity dispersion is 
$320 \pm 39$ ${\rm m}\, {\rm s}^{-1}$ and the distance modulus of the 
centroid of our sample of $43$ members is $(m-M) = 3.34 \pm 0.02$.
This distance modulus agrees with that determined by both P98 and PSSKH98 
and the velocity dispersion is consistent with the estimates of both P98
and D97. 
However, the difference between our estimate of the bulk velocity and that
of P98 arises from our more stringent selection of Hyades members.
\item [{(2)}:] The Hipparcos parallaxes of the three Hyades binary 
systems are consistent at the $1\sigma$ level with the parallaxes from 
their orbital solutions.
Hence, the systematic error in the Hipparcos parallaxes towards the Hyades
cluster is less than $0.47$ mas.
This is already more precise (by a factor of almost 1.4) than a 
straightforward comparison between the \h and orbital parallaxes performed
by P98, in which the precision is limited by the \h parallax errors, and hence
is irreducible in the future.
\item [{(3)}:] The test proposed in this paper can, in principle, 
 detect any systematic error greater than $0.3$ mas in the Hipparcos 
parallaxes towards the Hyades cluster.
The dominant factor that currently limits a check at this level is the 
``large'' errors in the orbital parallaxes of the binary systems.
\end{description}
It follows from the last two points that a more accurate estimate of the 
binary orbital parallaxes would enable a  better determination 
of the systematic errors (or the lack thereof) in the Hipparcos astrometry
towards the Hyades cluster.

\acknowledgments

This work was supported in part by the grant AST 97-27520 from the NSF.
We thank Marc Pinsonneault for providing the Hyades isochrones and 
for many helpful discussions.
We are grateful to Bob Hanson for his perspicuous comments on an earlier 
draft of this paper.
We also thank the referee for a detailed report that contained many useful
suggestions.



\end{document}